\documentclass[prl,floatfix,twocolumn]{revtex4-1}
\usepackage{graphicx} \usepackage{subfigure}
\usepackage{stdclsdv} \usepackage{array} \usepackage{amsmath}
\usepackage{amsfonts}

\usepackage{currfile}
\usepackage{fancyhdr}
\usepackage{datetime}
\newcommand{\Fig}[1]{Figure~\ref{#1}}
\newcommand{\Figs}[1]{Figures~\ref{#1}}
\newcommand{\Eq}[1]{Equation~\ref{#1}}
\newcommand{\Eqs}[1]{Equations~\ref{#1}}

\newcommand{\df}[0]{\Delta\! F}
\newcommand{\tn}[0]{\tau_{\mathrm{nucl}}}

\begin{document}

\title{Suppression of crystalline fluctuations by competing structures in a supercooled liquid}

\author{Pierre Ronceray}
\affiliation{Princeton Center for Theoretical Science, Princeton University, Princeton, NJ 08544, USA}

\author{Peter Harrowell} \affiliation{School of Chemistry,
University of Sydney, Sydney N.S.W. 2006, Australia}

\begin{abstract}
  We propose a geometrical characterization of amorphous liquid
  structures that suppress crystallization by competing locally with
  crystalline order. We introduce for this purpose the \emph{crystal
    affinity} of a liquid, a simple measure of its propensity to
  accumulate local crystalline structures on cooling. This quantity is
  explicitly related to the high temperature structural covariance
  between local fluctuations in crystal order and that of competing
  liquid structures: favouring a structure that, due to poor overlap
  properties, anticorrelates with crystalline order reduces the
  affinity of the liquid.  Using a lattice model of a liquid, we show
  that this quantity successfully predicts the tendency of a liquid to
  either accumulate or suppress local crystalline fluctuations with
  increasing supercooling.  We demonstrate that the crystal affinity
  correlates strongly with the crystal nucleation rate and the
  crystal-liquid interfacial free energy of the low-temperature
  liquid, making our theory a predictive tool to determine easily
  which amorphous structures enhance glass-forming ability.
\end{abstract}

\maketitle
\thispagestyle{fancy}

The ultimate fate of a liquid upon slow cooling -- whether it
crystallizes or arrests into an amorphous glass -- depends on how
easily the crystal can nucleate in the supercooled liquid.  It has
long been speculated that this kinetic stability with respect to
crystallization can be related to the geometrical properties of the
local structures which, at the molecular scale, are particularly
stable. More specifically, such favoured local structures that are
geometrically adverse to crystallinity -- for instance due to
non-crystalline symmetries -- would enhance the stability of the
liquid. This idea can be traced back to Frank's 1952
proposal~\cite{frank_supercooling_1952} that the stability of
icosahedral coordination shells in pure metallic liquids might impede
crystallization of close packed cubic crystals.  From this starting
point, a substantial literature has
developed~\cite{royall_role_2015,ma}, exploring a variety of
approaches to the nature and influence of liquid structure. This
includes the study of geometrical frustration in liquids~\cite{frust},
the descriptive study of the distribution of local coordination
structures in liquids via computer simulations~\cite{sims} and,
recently, nano-focused electron scattering~\cite{nano}, and the search
for correlation between specific local structures and the local
relaxation rates~\cite{dyn}.

In spite of this considerable activity, explicit evidence of the
essential thesis, that there is a correlation between liquid structure
and crystallization kinetics, remains sparse.  Taffs and
Royall~\cite{paddy2016} have reported on the crystallization of a hard
sphere liquid subjected to a bias that favours 5-fold common neighbour
coordination. They established a clear dependence of the reduced
crystallization time on the magnitude of the bias field, confirming
that the liquid structure does indeed influence the rate of
crystallization. The clear result of ref.~\cite{paddy2016} is
something of an exception: generally, any adjustment of the
Hamiltonian to vary the liquid structure also changes the stability of
the crystal structure, along with that of any polymorph favoured by
the perturbation. For example, Molinero et al~\cite{molinero} found
that adjusting the strength $\lambda$ of the 3-body contribution in a
model silicon potential resulted in a maximal glass forming ability
that coincided with the value of $\lambda$ corresponding to the
crossover in stable crystal phases. As this point coincides with the
maximal depression of the freezing point, it is difficult to
disentangle the role of liquid structure from that of crystal
stability. A similar issue has arisen in experimental efforts to
confirm the influence of liquid structure. Lee et al~\cite{lee}
studied the kinetics of crystal nucleation in a Ti-Zr-Ni alloy in
which the degree of local icosahedral order could be varied with
composition. Here the increase in liquid icosahedral order coincided
with stabilization of a quasicrystal whose nucleation rate was
significantly greater than that of the cubic crystal, effectively
concealing whatever influence the liquid structure had on the
crystallization of the latter crystal.

\begin{figure*}[t]
  \centering
  \includegraphics[width=\textwidth]{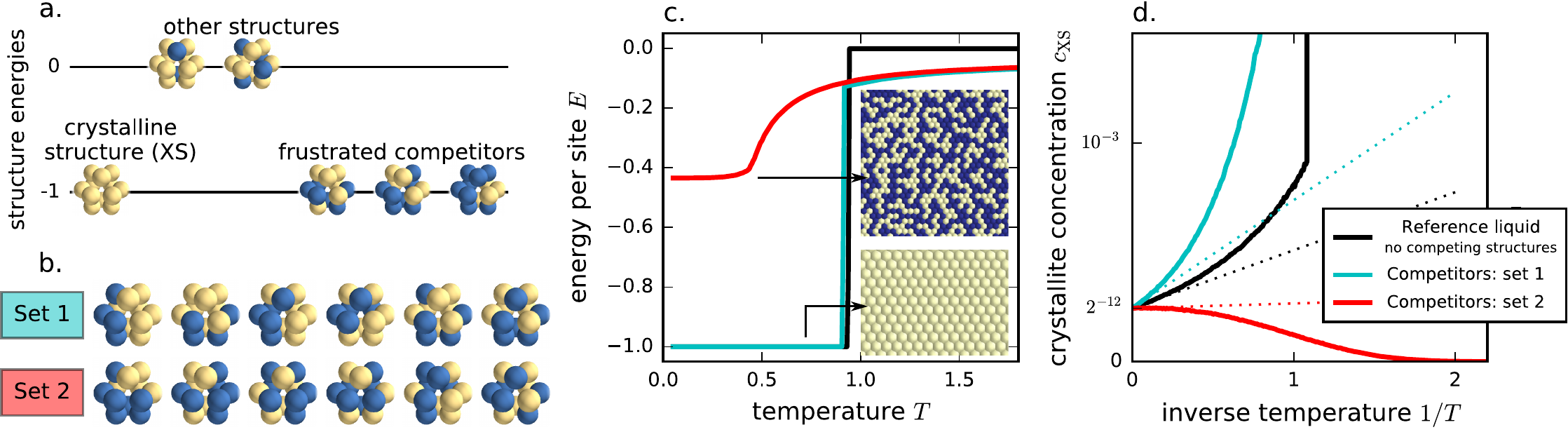}
  \caption{\textbf{a.} An illustration of the local energy landscape
    where the crystalline structure (XS) and a number $n$ of
    competitors have an energy of $-1$, while other structures have
    zero energy. \textbf{b.} Two sets of $n=6$
    competitors. \textbf{c.} Average energy per site as a function of
    temperature during a slow annealing (parameters: system size
    $30^3$, cooling rate $10^5$ MC steps / site / unit $T$). The Set 1
    liquid crystallizes faster than the reference liquid without
    competitors (lower inset). The Set 2 liquid does not crystallize,
    but arrests into an amorphous state (upper inset shows a
    slice). \textbf{d.} The concentration of the crystalline structure
    as a function of the inverse temperature for these three
    systems. Dotted lines indicate the leading order in $1/T$ with
    slope $Q$, the liquid's affinity to the crystalline structure, as
    predicted from Eq \ref{eq:c-T}. }
  \label{fig:Model}
\end{figure*}

In summary, to establish the influence of liquid structure on
crystallization kinetics we would like to be able to vary the liquid
structure as freely as possible while still ensuring that the
equilibrium crystal state remains unchanged.  The aim of this article
is to present both a model that satisfies these conditions, and a
theoretical framework relating the local geometrical properties of
stable liquid structures to the crystal nucleation rate. To this end,
we utilize a simple lattice model, the Favoured Local Structures (FLS)
model~\cite{ronceray_variety_2011,ronceray_2012,ronceray_favoured_2015}. We
consider binary spins, representing some local conformational degree
of freedom of the liquid (\emph{e.g.} composition), on a face-centered
cubic lattice. We define the \emph{local structure} of the liquid at a
given site as the geometrical arrangement of the 12 spins surrounding
it. There are 218 rotationally distinct structures at this
nearest-neighbor level, and we associate an energy $\epsilon_i$ to
each site with local structure $i$. Our model therefore has 218
variable energy levels corresponding to the possible local
structures. The set of energies $\{\epsilon_i\}_{i=1\dots 218}$
constitutes the \emph{local energy landscape} and entirely
characterizes the system's Hamiltonian. The energy per site is thus
\begin{equation}
  \label{eq:E}
  E = \sum_{\mathrm{structures\ }i} c_{i}\ \epsilon_i
\end{equation}
\noindent where $c_i$ is the fraction of sites in local structure
$i$. We study this model by canonical ensemble simulations using the
Monte-Carlo Metropolis algorithm with single spin flips, in a periodic
box with typical dimensions $30^3$.

In a previous study~\cite{ronceray_favoured_2015}, we have studied the
simplest case where a single $\epsilon_i=-1$ for a selected favoured
local structure, while all other structures have zero energy. In this
case, the energy per site is simply minus the concentration of the
FLS, and the ground state is the densest packing of the FLS. Because
structures at nearby sites overlap and thus exert constraints on each
other, these ground states can be highly frustrated (\emph{i.e.} the
selected FLS cannot fully ``tile'' the lattice) and the maximum FLS
fraction varies between $1/4$ and $1$.  As we have
shown~\cite{ronceray_favoured_2015}, this frustration does \emph{not}
prevent crystallization on moderately slow annealing of the system,
leading us to conclude that that some kind of \emph{competition}
between structures -- in other words, multiple minima in the local
energy landscape -- is required to stabilize the liquid. Indeed, we
have established more recently~\cite{ronceray_liquid_2016} that
favouring simultaneously two or more of the highly frustrated
structures can prevent crystallization even at very slow cooling
rates, resulting in the dynamic arrest of an amorphous state at low
temperature.

In this article, we use a variant of the FLS model adapted to study
the role of competing structures in the crystallization of a
supercooled liquid, as illustrated in \Fig{fig:Model}\textbf{a}. A
\emph{crystalline local structure} (denoted by ``XS'') competes with a
small number $n=2\dots 10$ of \emph{frustrated competing
  structures}. All these structures are given an energy
$\epsilon = -1$, while other structures have zero energy. For
simplicity, the crystalline structure is chosen to be the all-up
structure, whose ground state is unambiguously a uniform configuration
of up spins, with energy $-1$ per site. The mix of competitors is
selected such that it does not lead to the formation of alternate
metastable crystals, but continuously arrests (in the absence of the
XS) into a high energy amorphous state with high energy $E>-0.6$ (see
Methods).

An example demonstrates that the geometry of the frustrated
competitors can substantially influence the rate of crystallization of
the liquid. In \Fig{fig:Model}\textbf{b.} we present two apparently
similar sets of $n=6$ frustrated structures (Sets 1 and 2) that we
choose as competitors to the all-up crystalline structure. In
\Fig{fig:Model}\textbf{c.} we compare the behaviour of $E(T)$ of these
two systems on slow cooling with that of a reference liquid in which
only the crystalline structure is favoured. In this latter case (black
curve) the system exhibits little ordering in the liquid (less than
$0.1\%$ of XS) prior to a sharp first-order transition to the uniform
ground state (lower inset). Adding the Set 1 of competing structures,
the liquid is much more ordered and accumulates up to $12\%$ of FLS,
before freezing into the uniform ground state too. In contrast, Set 2
results in similar liquid energetics, but an absence of
crystallization: at low temperature, the system arrests into an
amorphous state with no trace of long-range ordering (upper inset of
\Fig{fig:Model}\textbf{c.}) in which $43\%$ of the sites are in a
favoured structure. We estimate the fastest nucleation rate of each of
these systems and find that the reference liquid (\emph{i.e.} that
with only the 'XS' favoured local structure) freezes in
$\tn \approx 7.10^4$ MC steps; with Set 1 favoured, freezing is
actually slightly faster with $\tn \approx 4.10^4$, while with Set 2
we find that $\tn > 10^{12}$ steps: we never observe crystallization
for this system in our simulations.

How can slight geometrical changes in the competing structures slow
down crystallization by more than seven decades? The origin of this
difference does not lie in the thermodynamic stability of the liquid:
they have comparable liquid energies and identical ground states. They
also have similar liquid relaxation times ($\tau_\alpha<10$ MC steps
per site). Their crucial difference is revealed by considering the
liquid structure. In \Fig{fig:Model}\textbf{d.} we plot the fraction
$c_{\mathrm{XS}}$ of sites in the crystalline structure for each of
these systems as a function of $1/T$. On cooling, we find that the Set
1 liquid accumulates significantly \emph{more} crystalline order than
the reference liquid: the competitors actually help the liquid
accumulating XS, a behaviour that we label as \emph{agonist} towards
crystallization. The Set 2 liquid, in contrast, \emph{antagonizes}
crystallization by suppressing local crystalline order: in spite of
the XS being favoured with energy $-1$, their concentration actually
decreases when cooling a liquid from the random infinite-temperature
limit. This observation intuitively explains the dramatic difference
in crystal nucleation rates in these two systems: crystalline local
structures are a necessary precursor to freezing, hence their
suppression can be expected to make crystallization extremely
difficult.

\begin{figure}[t]
  \centering
  \includegraphics[width=\columnwidth]{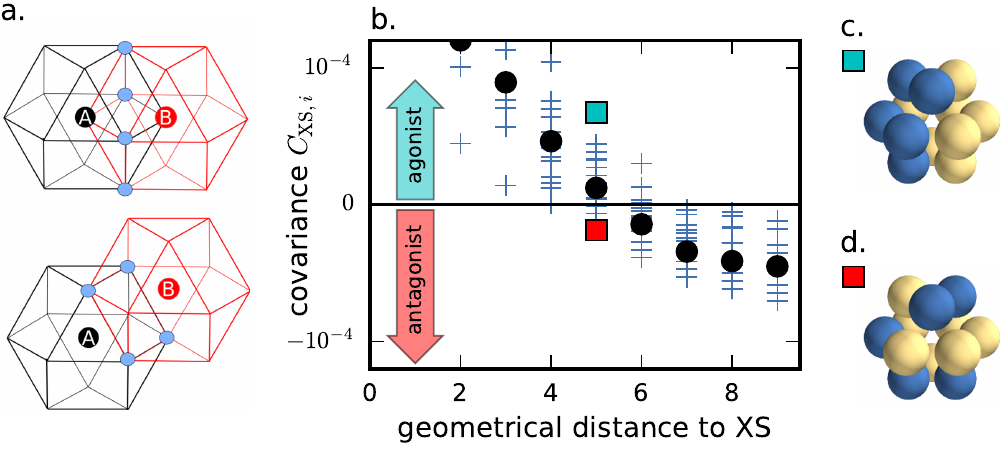}
  \caption{ \textbf{a.} Nearby structures on the lattice overlap and
    share sites (blue dots), giving rise to athermal correlations. \textbf{b.}  Structural covariance between the
    crystalline structure (all-up) and the 217 other structures
    (blue crosses), as a function of the number of spin flips required to
    convert this structure into the crystalline one. Black circles
    indicate average values.  \textbf{c.} and \textbf{d.}  Two structures with five
    down spins, the former being an agonist (cyan square in
    \textbf{b.})   while the
    latter is an antagonist (red square). }
  \label{fig:Cij}
\end{figure}

We now propose a general statistical measure of the liquid structure
that can quantitatively differentiate liquids like our Set 1 and Set 2
examples. A high-temperature expansion of the concentration
of local crystalline structures
\begin{equation}
  c_{\mathrm{XS}}(T)  = c_{\mathrm{XS},\infty}   + \frac{Q}{T} + O(T^{-2})
  \label{eq:c-T}
\end{equation}
already captures the distinction between crystal-agonist and
-antagonist systems through the value of the first-order coefficient
$Q$, that we call \emph{crystal affinity} of this liquid. Denoting by
$Q_0$ the affinity of the reference liquid with only the XS favoured,
agonist systems have $Q>Q_0$ and antagonists $Q<Q_0$ (dotted lines in
\Fig{fig:Model}\textbf{d}). In a previous
work~\cite{ronceray_liquid_2016}, we have developed an approach, that
we termed \emph{structural covariance}, to compute exactly such
high-temperature coefficients in terms of geometrical overlap of
structures. Indeed, we have shown the following fluctuation-response-like relation
\begin{equation}
  Q =  \left.  \frac{\partial c_{\mathrm{XS}} }{ \partial \beta}\right|_{T=\infty} =- \sum_{\mathrm{structures\ }i}  C_{\mathrm{XS},i} \  \epsilon_i\label{eq:Q}
\end{equation}
where $\beta = 1/T$, and the coefficients $C_{\mathrm{XS},i}$
quantifies the geometrical interactions between the XS and all
structures $i$ (including the XS itself). These coefficients are equal
to the covariances of structural concentrations,
\begin{equation}
  \label{eq:C}
  C_{i,j} = N\ \mathrm{Cov}_{T=\infty}(c_{i},c_{j})
\end{equation}
where $N$ is the system size (making $C_{i,j}$ size-independent), and
the covariance is computed at infinite temperature, \emph{i.e.} on
completely random spin configurations. In this regime, correlations
between structures only occur when structures overlap and share sites,
exerting athermal constraints on each other, as sketched in
\Fig{fig:Cij}\textbf{a}.  These covariances can thus be either
measured by statistical analysis of high-temperature liquid
configurations, or computed exactly by analyzing such
overlaps~\cite{ronceray_2012,ronceray_liquid_2016}. Intuitively,
similar structures will tend to overlap well and thus have a positive
covariance, while very different structures will tend to exclude one
from the other's vicinity. In the FLS model, we can define a
\emph{geometrical distance} between two structures as the minimum
number of spin flips required to convert one local structure into the
other. As shown in \Fig{fig:Cij}\textbf{b}, the value of
$C_{\mathrm{XS},i}$ decreases with geometrical distance between
competing structure $i$ and the XS, a trend that is not
specific to this choice of crystalline structure (Appendix). What is
also clear from \Fig{fig:Cij}\textbf{b} is that the geometrical
distance does not fully determine the structural
covariance. Structures in \Fig{fig:Cij}\textbf{c} and \textbf{d} both
are $5$ spins flips away from the XS, but the former is a strong
agonist while the latter is an antagonist (respectively cyan and red
squares in \Fig{fig:Cij}\textbf{b}). A glimpse at their geometry
explains this difference: the first has all its down spins on one
side, attracting XS on the other. The antagonist, in contrast, has its
down spins scattered over the structure, such that no XS can overlap
with it in any of the two ways shown in \Fig{fig:Cij}\textbf{a}. The
crystal affinity $Q$ defined in \Eq{eq:Q} therefore encodes, at a pair
interaction level, the specific geometric interactions between the
crystalline local structure and its frustrated competitors.

\begin{figure}[t]
  \centering
  \includegraphics[width=\columnwidth]{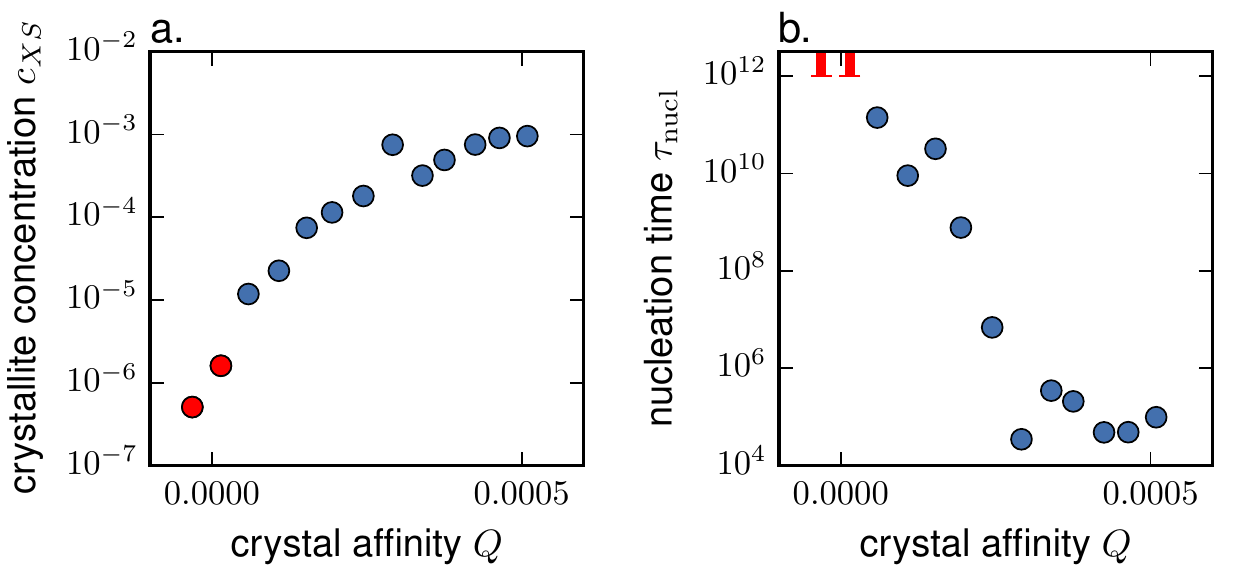}
  \caption{ \textbf{a.} Influence of the crystal affinity $Q$
    (\Eq{eq:Q}) of liquids with $n=7$ competitors on the concentration
    of XS, at low temperature $T=0.6$ where nucleation is the
    fastest. To prevent the liquid from crystallizing, the XS was not
    favoured in this specific plot. \textbf{b.} Fastest nucleation
    time of the crystal as a function of $Q$ (see Methods). The red
    error bar indicates no nucleation in $10^{12}$ MC steps. The 13
    systems presented here are selected to uniformly span the range of
    $Q$ values.}
  \label{fig:Qtau}
\end{figure}

The crystal affinity $Q$ is only a first-order term in \Eq{eq:c-T},
and so does not quantitatively capture the crystallite concentration
at low temperature in \Fig{fig:Model}\textbf{d}. It does very well,
however, in capturing both the degree of antagonism (or agonism) of
the liquid and the variation in different liquids in $c_{\mathrm{XS}}$
at low temperature, as shown in \Fig{fig:Qtau}\textbf{a}. This simple
geometrical quantity, therefore, provides a very useful insight into
the complex structural fluctuations of the supercooled liquid. Perhaps
most striking, the crystal affinity also correlates strongly with the
nucleation time of the crystal in this liquid. Indeed, in
\Fig{fig:Qtau}\textbf{b} we show that varying $Q$ in similar liquids
leads to variations of $\tn$ over eight orders of magnitude. We remind
the reader that $Q$ contains only information about structural pair
correlations obtained in the high temperature limit.

The mechanism by which the crystal affinity controls nucleation rates
can be interpreted within the framework of Classical Nucleation Theory
(CNT)~\cite{cnt}, in which the bulk free energy gain $\df$ in crystallization is
opposed by a surface free energy cost $\sigma$ for the interface
between the growing crystal nucleus and the supercooled liquid. In
this theory, the kinetically limiting step to crystallization is
assumed to be the thermally activated formation of a critical nucleus,
predicting an average nucleation time
\begin{equation}
  \tn = \tau_\alpha  N \exp\left( \frac{ \sigma^3}{k_B T \ \df^2}  \right)
  \label{eq:CNT}
\end{equation}
where $N$ is the number of potential nucleation sites and
$\tau_\alpha$ is the microscopic relaxation time of the supercooled
liquid. There are therefore three possible ways to make
crystallization slower: by slowing down the kinetics of the liquid
(\emph{i.e.} increasing $\tau_\alpha$), by stabilizing
thermodynamically the liquid (\emph{i.e.} lowering $\df$), or by
increasing the liquid-crystal surface tension $\sigma$.

\begin{figure}[t]
  \centering
  \includegraphics[width=\columnwidth]{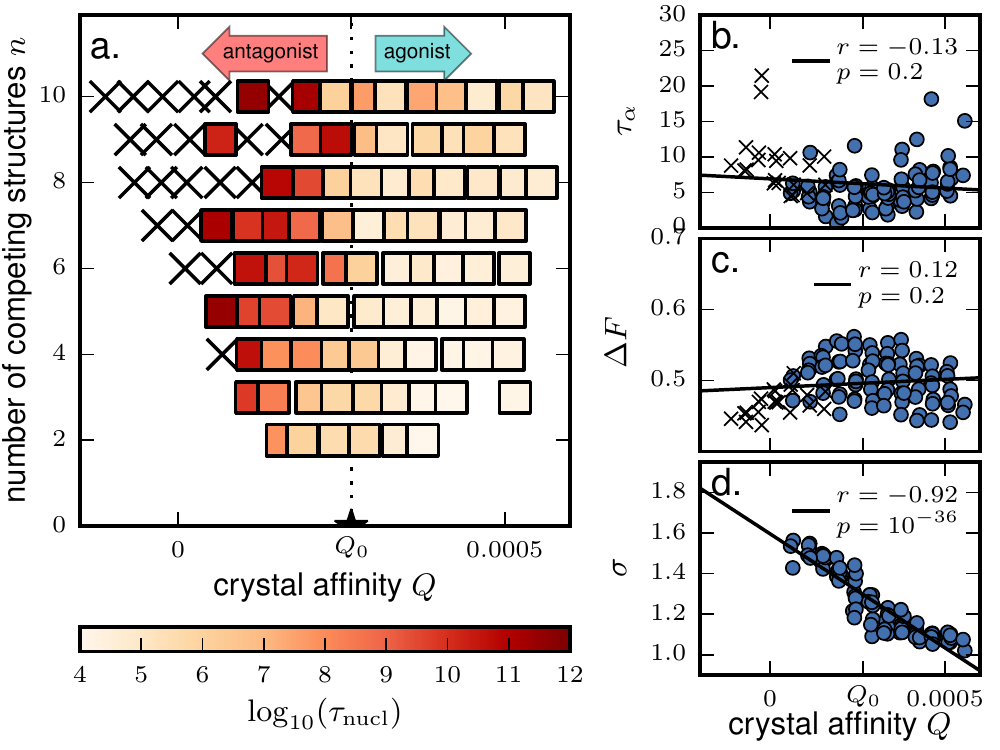}
  \caption{ \textbf{a.} Nucleation rates of liquids with varying
    crystal affinity $Q$ and number $n$ of frustrated competitors (see
    Methods), showing that it is the value of $Q$, rather than the
    number of competitors, that determines the crystallization
    rate. Black crosses indicate $\tn>10^{12}$ MC steps. \textbf{b-d.}
    Correlation between $Q$ and the three parameters of classical
    nucleation theory: liquid relaxation time $\tau_\alpha$ (computed
    as the autocorrelation time of the energy, in MC steps per site),
    liquid-crystal free energy difference $\df$ (computed by
    thermodynamic integration), and liquid-crystal surface tension
    $\sigma$ (inferred using \Eq{eq:CNT} and the measured nucleation
    times). The legend indicates the Pearson $r$ coefficient and
    $p$-value. The only statistically significant correlation is
    between $Q$ and $\sigma$.}
  \label{fig:CNT}
\end{figure}

We have seen in \Fig{fig:Qtau}\textbf{b} that the crystal affinity $Q$
strongly influences the nucleation rate. In order to understand which
of these aspects are affected by this affinity, we study nucleation in
a large number of supercooled liquids with $n=2$ to $10$ competing
structures, and selected to sample uniformly the available range of
$Q$ values. As shown in \Fig{fig:CNT}\textbf{a.}, the key factor
influencing nucleation times in this data set is the crystal affinity,
rather than the number of competing structures. In particular, it is
worth noting that for negative values of $Q$ (a situation indicating
that crystalline structures, in spite of being favoured, are linearly
suppressed in $1/T$ as the liquid is cooled down), crystallization is
robustly suppressed and never observed in our simulations. For this
whole data set, we analyze in \Fig{fig:CNT}\textbf{b-d.} the
correlations between $Q$ and each of the three aspects of the CNT
nucleation rate.  Strikingly, we find that $Q$ exhibits no significant
correlation with either the relaxation time $\tau_\alpha$
(\Fig{fig:CNT}\textbf{b.}) or the free energy difference $\df$
(\Fig{fig:CNT}\textbf{c.}). It does, however, correlate very strongly
with the surface tension $\sigma$ (\Fig{fig:CNT}\textbf{d.}): with an
$r$ coefficient of $-0.92$, the surface tension is almost fully
determined by $Q$. This effect is remarkable as it implies that,
within our model, the complex problem of estimating the
low-temperature surface tension $\sigma$ -- a quantity that is
somewhat ill-defined in CNT, as it is not a property at thermodynamic
equilibrium, and the concept of surface is ambiguous for microscopic
clusters -- can be bypassed by measuring the crystal affinity $Q$, a
simple, high-temperature structural quantity.

In this article, we have introduced the crystal affinity
$ Q = \left.  \partial c_{\mathrm{XS}} /\partial \beta
\right|_{T=\infty}$
as a measure of the propensity of the liquid to accumulate crystalline
order on cooling, using the local crystal structure concentration
$c_\mathrm{XS}$ as an inherent probe of crystallinity. This number
encapsulates, in an intelligible way, the geometrical interactions of
the crystal with non-crystalline stable local structures that compete
with crystalline order in the liquid. Indeed, we have shown with
\Eq{eq:Q} that, in a fluctuation-response-like relation, $Q$ can be
computed by analyzing the covariances in the number of crystalline and
non-crystalline local structures at high temperature. Furthermore, we
have demonstrated that $Q$ provides an excellent predictor of the
qualitative trends in rate of crystal nucleation: low affinity implies
slow nucleation. More precisely, we have established a clear
correlation between $Q$ and the effective crystal-liquid interfacial
energy (as obtained from the nucleation time data through the
assumption of classical nucleation). This result raises the
interesting prospect of a treatment of crystal nucleation in terms of
the statistics of structural fluctuations, rather than relying on the
awkward imposition of a macroscopic interface. Since the only
requirement for \Eqs{eq:c-T}-\ref{eq:C} to hold is that the
Hamiltonian of the system can be formulated in terms of a short-range
local energy landscape, as in \Eq{eq:E}, the analysis presented here
should provide a quite general framework for the systematic study of
structural fluctuations in off-lattice supercooled liquids, and the
influence of these fluctuations on the rate of crystal nucleation. In
particular, structural covariances could be used as a guide to
engineer local energy landscapes that kinetically facilitate or hinder
the formation of a target crystal structure.

\vspace{5mm}

\textbf{Acknowledgements.} PR is supported by the Prix des Jeunes
Chercheurs de la Fondation des Treilles. PH is supported by the
Australian Research Council.


\clearpage
\appendix

\section{Methods}

Here we provide the details of the rationale and procedures associated with the specific choices of model parameters used in the paper and the properties used to characterise structural fluctuations and nucleation rates. In a previous publication~\cite{ronceray_favoured_2015_SI}, we have presented a detailed description of the FLS model, including a comprehensive survey of the phase behaviour of the model in the cases where only a single structure is favoured. Readers interested in the model are referred to Ref. ~\cite{ronceray_favoured_2015_SI}.  

\paragraph{Calculating the crystal affinity.}

\begin{figure}[b]
  \centering
  \includegraphics[width=0.7\columnwidth]{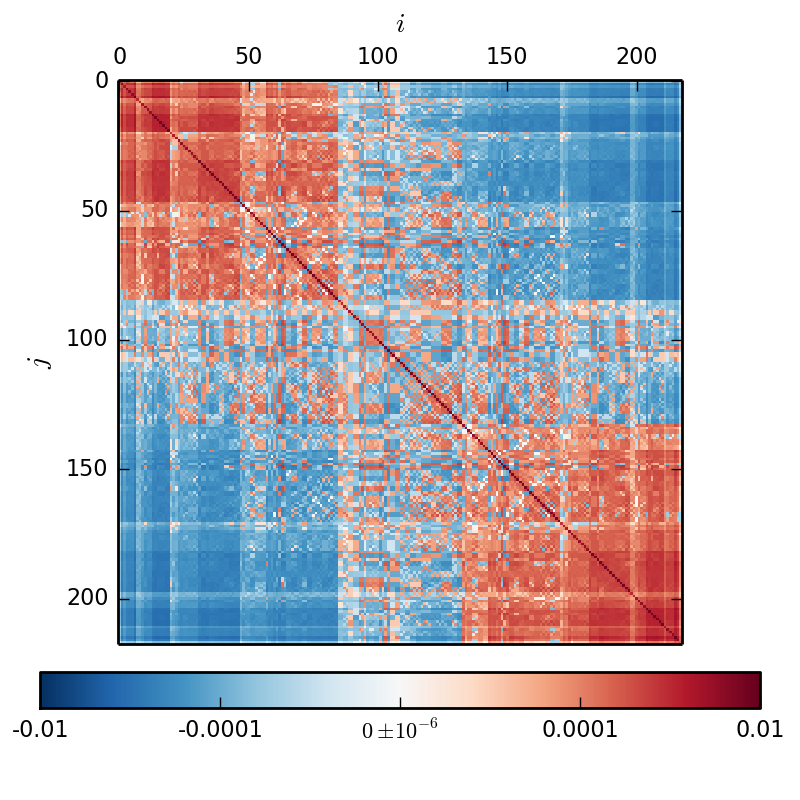}
  \caption{The covariance matrix of the FLS model at infinite
    temperature, showing all coefficients $C_{i,j}$ defined in
    \Eq{eq:C}. }
  \label{fig:CovMatrix}
\end{figure}

The high-temperature covariances between concentrations of local
structures, as introduced in \Eq{eq:C}, are displayed in
\Fig{fig:CovMatrix}. These coefficients are computed exactly using an
enumerative algorithm testing for all possible overlaps between pairs
of structures, as described in Ref.~\cite{ronceray_cluster_2016_SI,ronceray_liquid_2016_SI}.

\begin{figure}[t]
  \centering
  \includegraphics[width=0.7\columnwidth]{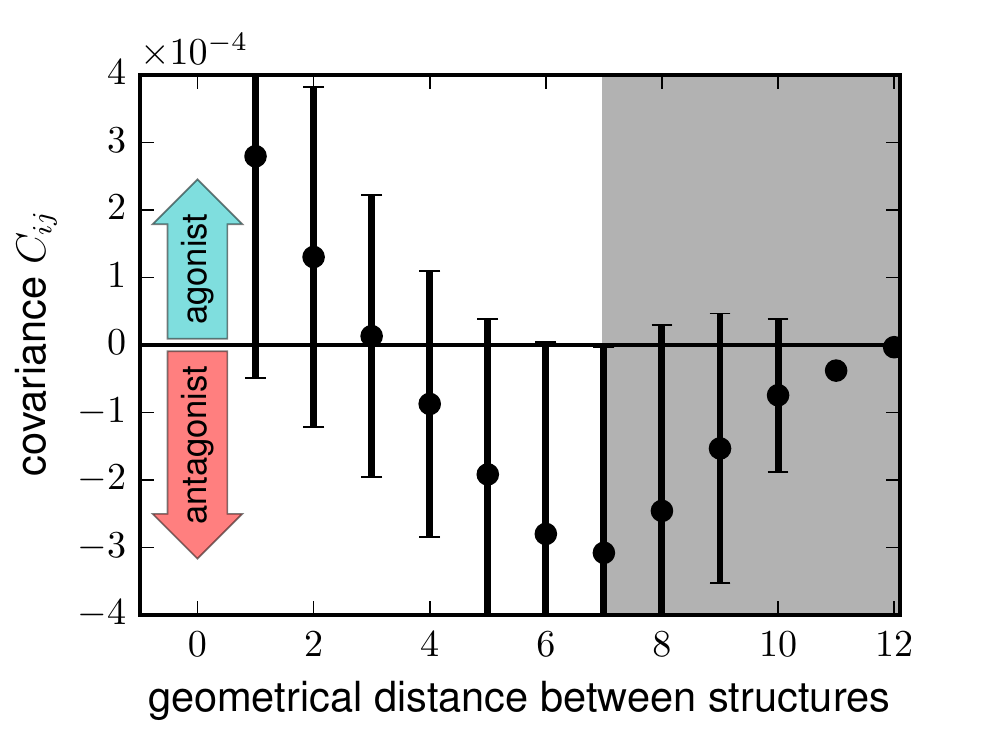}
  \caption{ The average covariance between any two structures as a
    function of their geometrical distance, complementing the plot in
    \Fig{fig:Cij}\textbf{b.} where only the all-up structure was
    shown. Error bars show standard deviation. The covarariances are
    clearly monotonically decreasing over the first 7 spin flips,
    though with large standard deviation reflecting the importance of
    the geometrical distribution of the spins. The increase for
    distance above 7 spin flips can be explained by symmetry effects
    (structures very far apart tend to have higher symmetry, which
    results in lower absolute value of the covariances as the mean
    concentrations are lower), and lack of data.}
  \label{fig:Cij_all}
\end{figure}

We show in \Fig{fig:Cij_all} that the coefficients $C_{i,j}$ depend on
geometrical distance between structures, regardless of the nature of
$i$ and $j$ (not only for the all-up XS selected in the article).

\begin{figure*}[t]
  \centering
  \includegraphics[width=\textwidth]{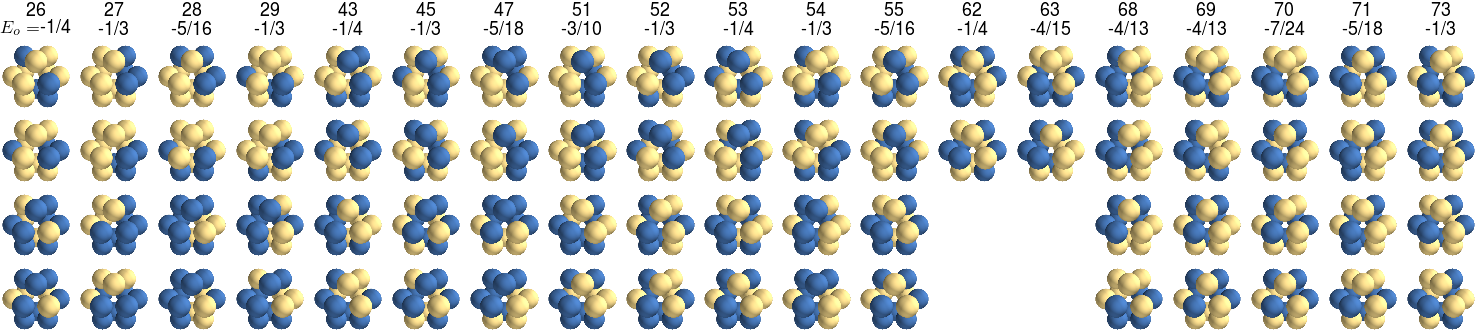}
  \caption{ The 19 choices of local structures used to generate the
    liquids studied in \Figs{fig:Qtau} and \ref{fig:CNT}, with their
    label as in Ref.~\cite{ronceray_favoured_2015_SI} and the
    corresponding ground state energy per site (minus the maximum
    fraction of sites that can be in the corresponding
    structure). Each column represents the spin- and mirror-inverted
    variants of the same structures, which on their own have identical
    properties, but have different packing properties when combined
    with other structures. Structures 62 and 63 have an additional
    symmetry (respectively spin inversion, and spin inversion combined
    with mirror symmetry) and thus have only two variants. Structures
    26-29 have 4 spins down; structures 43-55 have 5; and structures
    62-73 have six spins up and six down. Extensive details about
    these structures can be found in the Supplementary Information of
    Ref.~\cite{ronceray_favoured_2015_SI}. }
  \label{fig:structures}
\end{figure*}

\paragraph{Generating the liquids.}

We discuss here the way we generate and select the 107 liquids that
are used for the statistical study in \Fig{fig:CNT}. These are
selected randomly from a large set of possible combinations of the
$74$ structures depicted in \Fig{fig:structures}. These structures are
those of the $218$ local structures in the FLS model that have the
following properties:
\begin{itemize}
\item a ground state energy higher or equal than $-1/3$, ensuring that
  the structure is frustrated and does not permit formation of crystal
  polymorphs that would compete with the all-up ground state;
\item a number of ``up'' spins between $4$ and $8$, so that the
  interactions between structures rely on geometrical properties
  rather than on mere ferromagnetic attraction or repulsion.
\end{itemize}
There are $19$ distinct types of local structures (columns in
\Fig{fig:structures}) fulfilling these conditions, coming in up to
four variants when including the mirror- and/or spin-inverted
variants.  For each value of the number of competing local structures
$n=2...10$, we generate $10^4$ different Hamiltonians by picking
randomly one structure in $n$ randomly selected columns of
\Fig{fig:structures} (We don't use two structures of the same column
as we observed empirically that this tends to increase crystallization
of composite metastable crystals.). We then pick samples that are
regularly spaced in values of $Q$, so as to investigate the
correlation between $Q$ and thermodynamic observables. These samples
are selected by binning the $10^4$ systems by values of $Q$, with
interval $1.5 \times 10^{-5}$.  We then pick one system in one bin out
of three, thus obtaining a first batch of 107 Hamiltonians (not all
values of $(Q,n)$ are possible).

We are interested in nucleation of the target crystal structure (the
all-up system) within a liquid where the competitors are present
too. We thus run a pre-selection simulated annealing, with only the
frustrated competing structures favoured (no XS), to eliminate those
of these systems that tend to crystallize into a polymorph crystal
made of a combination of the frustrated competing structures, and thus
wouldn't remain liquid at temperatures near the point at which we
study nucleation, $T = 0.6$. We thus eliminate and re-pick the system
to obtain another Hamiltonian with similar $Q$ value, if, on slow
annealing:
\begin{itemize}
\item the system exhibit a sharp first-order phase transition apparent
  as a discontinuity in energy per site $E(T)$; or
\item the system exhibits a peak of heat capacity at $T>0.45$ (sign of
  either a weakly first order phase transition, or a second order
  transition that would result in (near-)critical slowdown of the
  kinetics, and hence affect the nucleation rate in an undesired way; or
\item the system exhibits obvious long-range crystalline ordering in
  the final, low-temperature state.
\end{itemize}
Once these pre-selection tests and repicking are performed, we have
obtained a large set of 107 Hamiltonians composed of a mixture of the
structures depicted in \Fig{fig:structures}, regularly spaced in $Q$
and $n$ values, that do not, when only these structures are favoured,
crystallize on cooling. This phenomenology is further described in
Ref.\cite{ronceray_liquid_2016_SI} in the case of mixtures of $2$
frustrated FLS's. Note that the $13$ systems presented in
\Fig{fig:Qtau} are simply the $n=7$ line in \Fig{fig:CNT}.

\paragraph{Measuring nucleation times.}

We can next measure nucleation times for each of these
Hamiltonians. Our protocol is the following: we prepare the liquid in
a system of $20^3$ sites with periodic boundary conditions, by
annealing a similar system, but without favouring the XS so as to
prevent premature nucleation. Having such an equilibrated liquid, we
check that it hasn't crystallized into a frustrated polymorph
(\emph{i.e.} that the system's energy is consistent with the
corresponding liquid), then add the XS to the set of favoured
structures. We then perform MC moves at a fixed temperature $T=0.6$
(chosen to be the temperature of fastest nucleation, in a quite robust
way) until crystallization is detected, \emph{i.e.} when the energy
falls below a threshold energy of $-0.9$, from which the system always
falls into the ground state. This protocol is however spoiled by the
presence of ballistic relaxation, both before nucleation (relaxation
of the liquid after adding the XS) and after nucleation (crystal
growth). For short nucleation times, this makes the measured time
before hitting the threshold strongly reproducible. We thus obtain an
upper bound to the nucleation time by repeating the simulation several
times (up to 20 times or $10^{12}$ MC steps) and measuring the
standard deviation of the crystallization time, which gives us an
estimate of $\tn$, assuming that nucleation times are exponentially
distributed (\emph{i.e.} no aging or memory in the liquid prior to
nucleation).

\paragraph{Liquid relaxation times.}

The relaxation times $\tau_\alpha$ of the liquid (as presented in
\Fig{fig:CNT}\textbf{b.})  are the autocorrelation times of the energy
of the liquid, expressed in Monte-Carlo steps per site. They are
obtained using the batch autocorrelation method.

\paragraph{Free energies.}

Liquid-crystal free energy differences presented in
\Fig{fig:CNT}\textbf{c.} are computed by approximating the crystal's
free energy by its energy value $E\approx -1$, neglecting its
entropy. An approximation of the liquid free energy is obtained by
thermodynamic integration from infinite temperature of the system
without the XS favoured, so that it does not crystallize in the
annealing run performed to integrate the free energy. Neglecting the
XS leads to a minor error in the free energy of the liquid (less than
$1\%$ in systems for which crystallization is slow enough to allow
measurement of the free energy with the XS favoured).

\begin{figure}[htb]
  \centering
  \includegraphics[width=0.7\columnwidth]{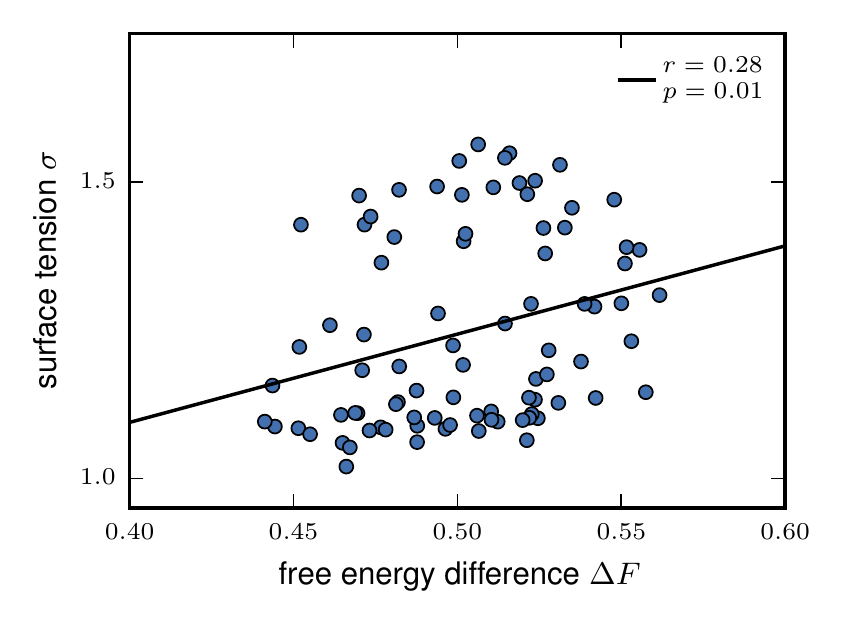}
  \caption{ Correlations between the liquid-crystal surface tension
    and bulk free energy difference are negligible within our data
    set. }
  \label{fig:sigma_vs_dF}
\end{figure}

\paragraph{Surface tensions.}

We do not measure directly the surface tensions presented in
\Fig{fig:CNT}\textbf{d.}; rather, we infer it through Classical
Nucleation Theory assumption, using
\begin{equation}
  \label{eq:CNT-sigma}
  \sigma \approx \left[ k_B T \df^2 \log( \tn / \tau_\alpha ) \right]^{1/3}
\end{equation}
where $k_B = 1$ in our unit system, $T=0.6$ for all nucleation runs
presented in this article, and $\tn$, $\tau_\alpha$ and $\df$ are
obtained as discussed above. Note that we do not include the volume
factor $N=20^3$ in \Eq{eq:CNT-sigma}; doing so would result in undefined
values for $\sigma$ for many of the high-affinity liquids, as
nucleation is often faster than $N \tau_\alpha$ in these cases. In any
case, $\sigma$ in this article should not be interpreted too strictly
as a surface tension -- as discussed at the end of the main text, this
notion is somewhat ill-defined, on top of being particularly difficult
to measure directly. Rather, it quantifies the aspect of the
nucleation rate that is not a consequence of the bulk free energy gain
$\df$ in crystallizing, as illustrated in \Fig{fig:sigma_vs_dF} by
showing the near absence of correlation between $\sigma$ and $\df$.

\begin{figure}[t!]
  \centering
  \includegraphics[width=\columnwidth]{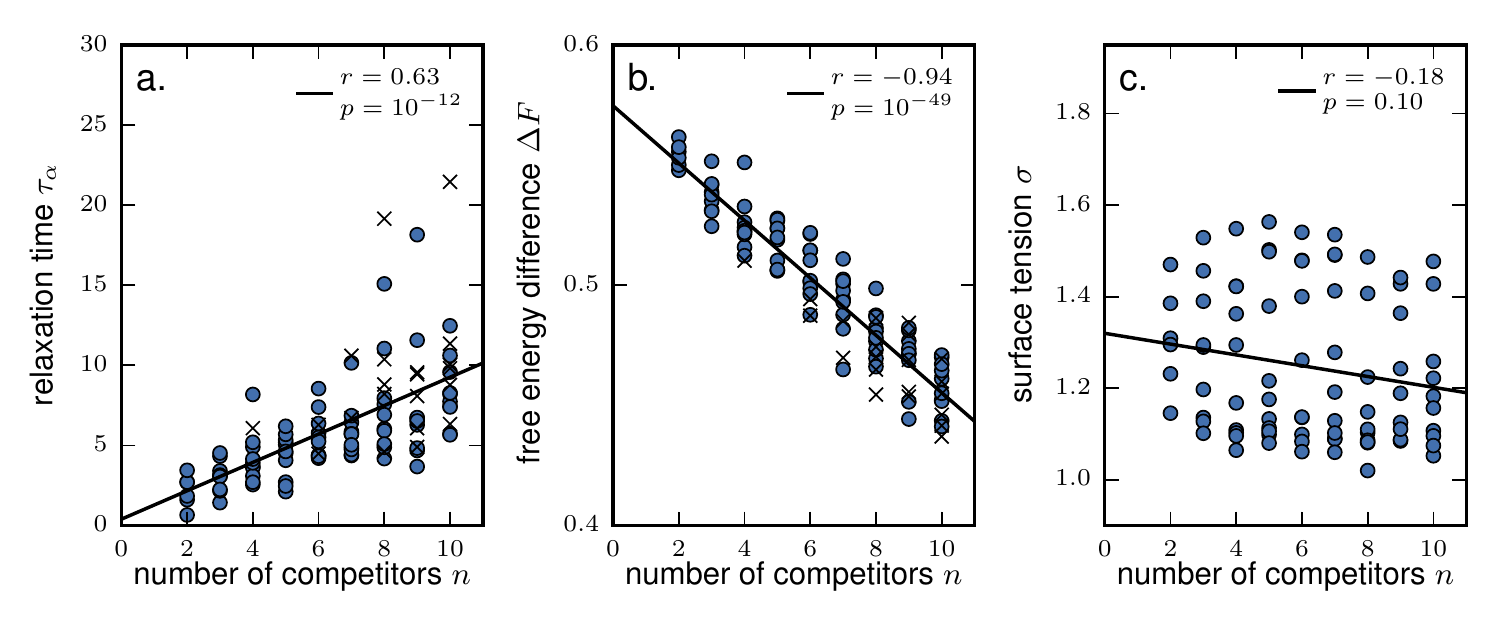}
  \caption{ Correlations between number of competing structures $n$
    and parameters of the Classical Nucleation Theory that influence
    the nucleation time. The data set is the same as in
    \Fig{fig:CNT}. }
  \label{fig:Ncorr}
\end{figure}

\paragraph{Influence of the number of competitors.}

In \Fig{fig:CNT}, we present a large data set of liquids with varying
number of competing structures $n$ and crystal affinity $Q$. Panels
\textbf{b-d.} of this figure analyze the influence of $Q$ on the
different parameters $\tau_\alpha$, $\df$ and $\sigma$ that affect the
nucleation time, showing that $Q$ strongly correlates with $\sigma$
while being essentially independent of the other parameters. As a
complement, we present in \Fig{fig:Ncorr} the same correlations, but
with the other parameter of our data set, the number of frustrated
competing structures $n$. It presents a strong correlation with $\df$
($r=0.94$), while showing no significant correlation to
$\sigma$. These results can be intuitively understood by noting that
adding more competitors to the liquid increases the number of ways for
the liquid to lower its energy: it will thus increase its entropy, at
a given energy, and thus lower the free energy of the liquid, while
leaving the crystal unaffected. On the other hand, these additional
competitors are equally likely to act as agonist or antagonist with
the crystalline structure, hence they do not significantly affect the
crystal affinity $Q$ -- and, therefore, do not affect the surface
tension either.

\end{document}